\documentclass[english]{aipproc}
\usepackage[T1]{fontenc}
\usepackage[latin1]{inputenc}
\usepackage{graphicx}

\makeatletter

 \usepackage{aipxfm}
 \layoutstyle{6x9}

\usepackage{babel}
\makeatother
\begin{document}
\title{The microscopic theory of fission}



\author{W. Younes}{address = {Lawrence Livermore National Laboratory, Livermore, CA 94551}}

\author{D. Gogny}{address = {Lawrence Livermore National Laboratory, Livermore, CA 94551}}

\begin{abstract}
Fission-fragment properties have been calculated for thermal neutron-induced
fission on a $^{239}\textrm{Pu}$ target, using constrained Hartree-Fock-Bogoliubov
calculations with a finite-range effective interaction. A quantitative
criterion based on the interaction energy between the nascent fragments
is introduced to define the scission configurations. The validity
of this criterion is benchmarked against experimental measurements
of the kinetic energies and of multiplicities of neutrons emitted
by the fragments.
\end{abstract}
\maketitle

\section{Introduction}

The description of fission as a quantum many-body problem is simultaneously
the most promising and the most difficult path toward a predictive
theory of this phenomenon. A full treatment of all possible many-body
configurations of the fissioning system is computationally unfeasible
and, in many cases, unnecessary. In practice, it is known that the
nucleus in its lowest-energy state is well described by a single Slater
determinant. Thus, mean-field approaches such as the Hartree-Fock
Bogoliubov (HFB) theory have been extremely successful in describing
the fission process \cite{berger84,warda02,goutte05,dubray08}.

A microscopic fission-theory program is being developed at the Lawrence
Livermore National Laboratory which describes the fissioning system
in terms of its constituent protons, neutrons, and the effective (i.e.,
in-medium) interaction between nucleons. This approach is based on
the highly successful program developed at the Bruyères-le-Châtel
laboratory over the last three decades \cite{berger84,goutte05,dubray08},
and provides a fully microscopic, quantum-mechanical, dynamical, and
self-consistent description of fission. The only phenomenological
input to the method is the effective interaction between nucleons,
and the D1S finite-range interaction \cite{decharge80} has been used
in this work.

In the first phase of the program, devoted to the static aspects of
fission, we have focused on the definition and analysis of scission
configurations, where the nucleus divides into (typically two) distinct
fragments. The HFB formalism is the main tool used in this analysis.
Static HFB calculations can be performed for specific configurations
of the nucleus through the use of constraints on various collective
{}``coordinates'' of the nucleus (e.g., quadrupole/octupole/hexadecapole
moments, number of particles in the neck, separation distance between
fragments). Among these configurations, some will correspond to a
single whole nucleus, while others will describe two distinct fragments.
In this work, we will be interested in the boundary between these
two regions in configuration space, and the properties of the nascent
fragments (shape, kinetic and excitation energies) that can be extracted
from the calculations.

\section{Theory}

Detailed descriptions of the HFB formalism with constraints can be
found in the literature \cite{decharge80,younes09}. Here, we only
recall the salient point of the theory. The lowest-energy state of
the fissioning system characterized by a Hamiltonian $\hat{H}$ and
a set $\left\{ q_{i}\right\} $ of collective coordinates is found
by the variational principle\begin{eqnarray*}
\delta\left\langle \left\{ q_{i}\right\} \left|\hat{H}-\lambda_{N}\hat{N}-\lambda_{Z}\hat{Z}-\sum_{i}\lambda_{i}\hat{Q}_{i}\right|\left\{ q_{i}\right\} \right\rangle  & = & 0\end{eqnarray*}
subject to the constraints\begin{eqnarray*}
\left\langle \left\{ q_{i}\right\} \left|\hat{N}\right|\left\{ q_{i}\right\} \right\rangle  & = & N\\
\left\langle \left\{ q_{i}\right\} \left|\hat{Z}\right|\left\{ q_{i}\right\} \right\rangle  & = & Z\\
\left\langle \left\{ q_{i}\right\} \left|\hat{Q}_{i}\right|\left\{ q_{i}\right\} \right\rangle  & = & Q_{i}\end{eqnarray*}
where $N$ and $Z$ are the neutron and proton numbers, respectively,
and $Q_{i}$ is any of the remaining collective-coordinate values.
These collective-coordinate values are calculated as the expectation
value of corresponding operators $\hat{N}$, $\hat{Z}$, and $\hat{Q}_{i}$.
The $\hat{Q}_{i}$ are typically multipole operators, but we have
also used the neck-size constraint\begin{eqnarray*}
\hat{Q}_{N} & \equiv & \exp\left[-\frac{\left(z-z_{N}\right)^{2}}{a_{N}^{2}}\right]\end{eqnarray*}
where $z$ is measured along the symmetry axis of the nucleus, $z_{N}$
is the position of the neck (i.e., where the neck is thinnest), and
$a_{N}=1\,\textrm{fm}^{2}$. The many-body Hamiltonian $\hat{H}$
is expressed in terms of an effective finite-range density-dependent
interaction between the nucleons with the D1S parameterization \cite{decharge80}.
The D1S parameters were adjusted to properties of $^{16}\textrm{O}$,
$^{90}\textrm{Zr}$, Sn isotopes, and infinite nuclear matter. The
only fission-related constraint on the interaction was introduced
through a slight readjustment of the surface-energy term in nuclear
matter to better reproduce the height of the fission barrier in $^{240}\textrm{Pu}$.

We have implemented the constrained HFB formalism with finite-range
effective interaction in a code that uses a one-center axially deformed
harmonic-oscillator basis. In this work we assume axial symmetry,
and the Hamiltonian matrix assumes a block-diagonal form, with the
blocks labeled by the angular-momentum projection quantum number $\Omega$.
The present calculations have been performed with up to 27 harmonic-oscillator
shells in the axial direction. The two-body center-of-mass correction
has been included. The Slater approximation has been used for the
Coulomb-exchange interaction. Only the central part of the effective
interaction has been included in the pairing interaction.

\section{Results}

\begin{figure}
\includegraphics[%
  scale=0.44,
  angle=-90]{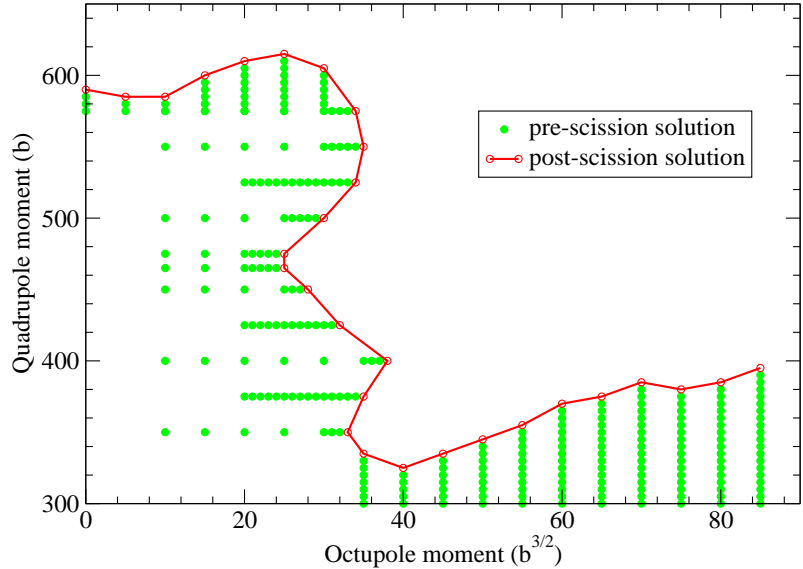}

\caption{\label{cap:240pu-scl23-new}Scission line for $^{240}\textrm{Pu}$
hot fission obtained in this work. The solid green points represent
HFB calculations producing a whole (non-scissioned) nuclear density.
The empty red circles connected by a solid line represent scissioned
configurations.}
\end{figure}

In this work, we have explored two distinct criteria for the definition
of scission. The first definition is based on the value of $Q_{N}$,
the second on the interaction energy between the fragments. The scission
configurations identified with a $Q_{N}$-based criterion are shown
in Fig. \ref{cap:240pu-scl23-new} for $^{240}\textrm{Pu}$ fission.

The figure shows lines of HFB calculations at fixed $Q_{20}$ or $Q_{30}$
values, each using the previous solution as a starting point. The
scission line was identified by a drop in $Q_{N}$ to relatively small
values (i.e., $Q_{N}\ll0.5$) along a given line of calculations.
The scission configurations identified in this manner correspond to
the {}``hot fission'' mode \cite{berger84}. In this mode, the fragments
are formed relatively far apart and therefore with comparatively reduced
kinetic energies, and correspondingly higher excitation energies.
This fission mode is expected to dominate low-energy induced fission,
such as in the $^{239}\textrm{Pu}\left(\textrm{n}_{\textrm{th}},\textrm{f}\right)$
reaction.

The scission line obtained in Fig. \ref{cap:240pu-scl23-new} represents
a non-trivial boundary separating regions where the nucleus is either
whole or scissioned. Similarly-complicated boundaries have been previously
observed by Dubray et al. \cite{dubray08} in their studies of Th
and Fm fission. The behavior of the nucleus as it crosses the scission
line is different near mass-symmetric ($Q_{30}=0$) and most probable
($Q_{30}=60\,\textrm{b}^{3/2}$) fission limits. Near the symmetric
limit, the variation in calculated properties (neck size, total HFB,
energy, etc.) between points just before and just after the scission
line is far greater than in the asymmetric case. These large variations
make it impossible to extract the properties of the fragments at scission
using the quadrupole and octupole constraints alone. Other constraints
can be introduced, such as the hexadecapole moment $Q_{40}$ and the
neck parameter $Q_{N}$. In our analysis, we have found $Q_{N}$ to
be a more effective constraint to hold the nucleus back from scission,
but harder to control than $Q_{40}$. In addition the $Q_{40}$ and
$Q_{N}$ constraints are not simply related: for a given $Q_{40}$
value, it is possible to find distinct configurations with either
a significant neck or a vanishing one \cite{younes09}. For the results
shown in this paper, we have used the $Q_{N}$ constraint to approach
the scission configurations in a more controlled manner. In the future,
we plan to use both $Q_{N}$ and the distance between fragments to
analyze the detailed behavior of the nucleus near scission.

\begin{figure}
\includegraphics[%
  scale=0.44,
  angle=-90]{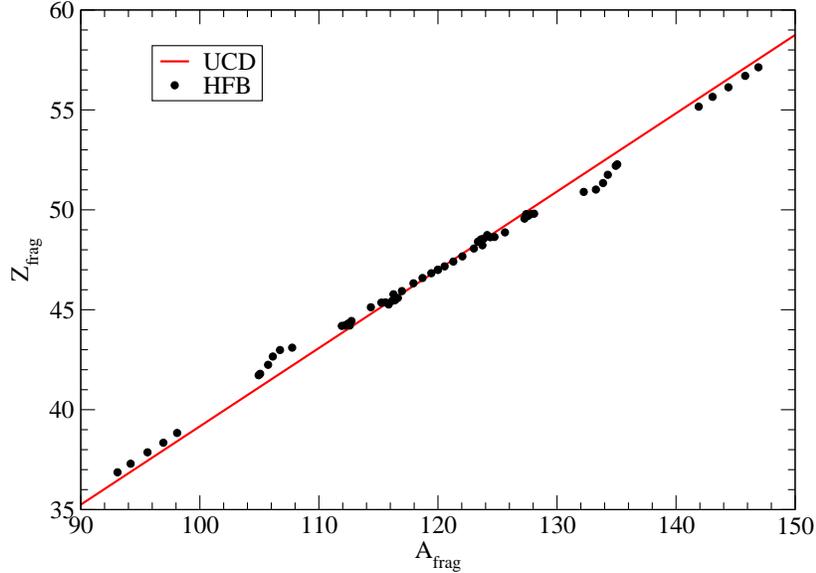}

\caption{\label{cap:240pu-frag-z-curvy}Fission-fragment charge number plotted
as a function of mass number obtained from HFB calculations just before
scission in Fig. \ref{cap:240pu-scl23-new}. The UCD prediction (solid
red line) is plotted for comparison.}
\end{figure}

Using the value of $Q_{N}$ as an indicator of scission, we have examined
the shape properties of the nascent fragments at a point close to
scission but such that the nucleus is still whole. In practice, these
properties were extracted at the last (green) point along each line
of calculations in Fig. \ref{cap:240pu-scl23-new} before the (red)
scission line. In Fig. \ref{cap:240pu-frag-z-curvy}, we show the
HFB calculations of the charge and mass numbers of the fragments at
the last pre-scission point, compared to the prediction of the Unchanged
Charge Distribution (UCD) model \cite{wahl88}. These particle numbers
were obtained by integrating the proton and neutron densities to the
left and right of the neck position $z_{N}$ along the symmetry axis
of the nucleus. The HFB and UCD predictions are in excellent agreement.
This result is even more remarkable when one realizes that the only
phenomenological ingredient in the HFB calculation, i.e. the effective
interaction, was not explicitly adjusted to reproduce this property.
Similarly, the quadrupole moment of the fragments has been extracted
from the HFB calculations and, although the calculations exhibit some
fluctuations, a trend emerges. As expected, the $Q_{20}$ values drop
significantly near the A = 134 mass, which is dominated by the near-spherical
$^{134}\textrm{Te}$ fragment.

Despite its usefulness, the $Q_{N}$ criterion for scission does present
some difficulties in its interpretation. As note before, the properties
of the nucleus immediately before and after scission generally differ
greatly, especially in the case of symmetric fission. Therefore, it
is not clear that the kinetic and excitation energies of the fragments,
which are extremely sensitive to the value of $Q_{N}$, can be correctly
reproduced when adopting a criterion based on neck size. The difficulty
arises because there is no objective quantitative criterion for the
neck size at scission. As an alternative, we consider a definition
of scission based on the interaction energy between the fragments.
In order to formulate a quantitative criterion we adopt the ansatz
that, in the present static calculations with constrained $Q_{N}$,
scission occurs with decreasing $Q_{N}$ as soon as there is enough
energy available in the system to overcome the attractive part of
the interaction between the fragments.

For each final (green) pre-scission point in Fig. \ref{cap:240pu-scl23-new},
the $Q_{N}$ value is progressively decreased in small increments
(typically $\Delta Q_{N}=0.05$). For each HFB calculation at constrained
$Q_{N}$ value, the single-particle wave functions are classified
as either predominantly localized to the left or to the right of the
neck position $z_{N}$. At this stage, the total particle densities
for the left and right fragments can be readily calculated from their
respective single-particle wave functions. The density for each fragment
will typically exhibit a tail that extends into the complementary
fragment. Often these tails can be quite large containing from a few
to over a hundred nucleons in the asymmetric- and symmetric-fission
limits, respectively. These tails can be reduced by a change in representation
before calculating the fragment properties. In practice, for each
pair of single-particle wave functions $\psi_{i}$ and $\psi_{j}$
within the same symmetry block in the density matrix, an angle $\theta$
is sought such that the number of particles in the tails of the rotated
wave functions\begin{eqnarray*}
\left(\begin{array}{c}
\psi_{i}^{\prime}\\
\psi_{j}^{\prime}\end{array}\right) & = & \left(\begin{array}{cc}
\cos\theta & -\sin\theta\\
\sin\theta & \cos\theta\end{array}\right)\left(\begin{array}{c}
\psi_{i}\\
\psi_{j}\end{array}\right)\end{eqnarray*}
is reduced. This transformation will not yield a reduction in the
tails for all possible pairs of wave functions $\left(\psi_{i},\psi_{j}\right)$,
and it is only applied to those pairs where such a reduction is achieved.
All possible pair combinations are processed in this manner, and the
entire procedure is iterated to suppress the tails even further. In
practice, 30 iterations were sufficient to reach a decrease in tail
size of less than 0.5\% between subsequent iterations.

Working in this reduced-tail representation, it is now possible to
calculate the interaction energy between fragments defined as\begin{eqnarray*}
E_{int} & = & E_{HFB}-E_{HFB}\left(L\right)-E_{HFB}\left(R\right)-E_{coul}^{\left(D\right)}\end{eqnarray*}
where $E_{HFB}$ is the HFB energy for the entire fissioning nucleus,
$E_{HFB}\left(L\right)$ and $E_{HFB}\left(R\right)$ are the HFB
energies of the left and right fragment respectively (calculated using
the generalized densities identified for each fragment), and $E_{coul}^{\left(D\right)}$
is the direct contribution of the Coulomb energy between the fragments.
Care was taken to use the density of the whole nucleus (and not that
of the left or right fragment) when calculating the contribution to
$E_{HFB}\left(L\right)$ and $E_{HFB}\left(R\right)$ from the density-dependent
part of the interaction. The direct Coulomb term was calculated by
integrating the product of fragment proton densities, folded with
an inverse-distance potential. The energy $E_{int}$ therefore contains
contributions from the purely-nuclear potential as well as the attractive
Coulomb-exchange term between fragments.

For low-energy fission, we assume that the energy available in the
system is measured from the top of the second barrier. If $\left\{ q\right\} $
denotes an arbitrary set of constraints, and $\left\{ q_{B_{II}}\right\} $
the set of constraints at the second barrier, we measure the energy
available in the system as $\Delta E=E_{HFB}\left(\left\{ q\right\} \right)-E_{HFB}\left(\left\{ q_{B_{II}}\right\} \right)$.
If scission is approached by varying a specific constraint, say the
neck size $Q_{N}$, we then identify the scission configuration as
the first instance where $\Delta E=E_{int}$. This configuration corresponds
to our ansatz of building up enough energy in the system to overcome
the attractive part of the interaction potential between the fragments.

\begin{figure}
\includegraphics[%
  scale=0.44,
  angle=-90]{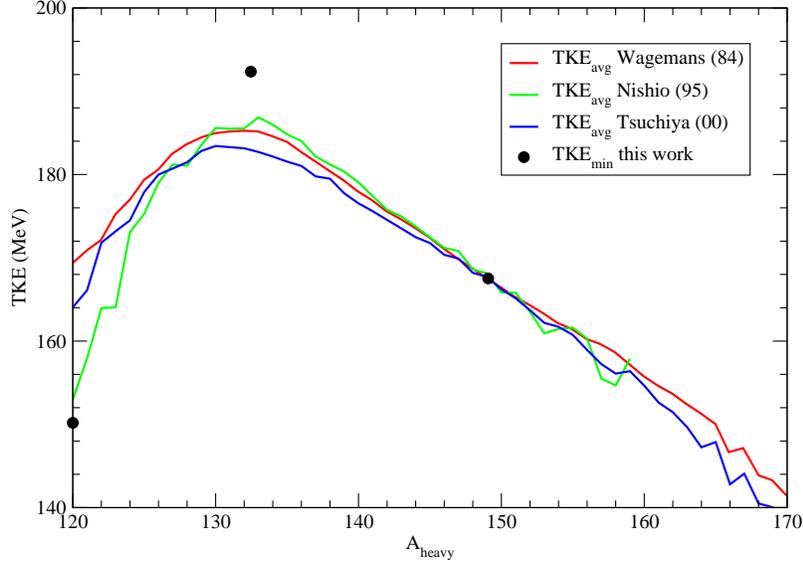}

\caption{\label{cap:tke-a}Comparison between measured \cite{wagemans84,nishio95,tsuchiya00}
and calculated total kinetic energies of the fragments, plotted as
a function of the heavy-fragment mass.}
\end{figure}

Using this interaction-energy based criterion, we have calculated
the kinetic and excitation energies of the fragments for a few selected
points along the scission line in Fig. \ref{cap:240pu-scl23-new}.
In particular, we have made these calculations for the symmetric ($Q_{30}=0$),
most probable ($Q_{30}=60\,\textrm{b}^{3/2}$), and very asymmetric
($Q_{30}=85\,\textrm{b}^{3/2}$) fission limits. The calculated Total
Kinetic Energies (TKE) are compared in Fig. \ref{cap:tke-a} to three
experimental measurements \cite{wagemans84,nishio95,tsuchiya00}.
The experimental data agree everywhere, except in the symmetric-fission
limit, where the measured TKE vary widely from $\approx$ 153 to 169
MeV. In the symmetric limit, the HFB calculations give a value for
the TKE that is closest to the Nishio result, but in general, the
calculated values agree with all three measurements everywhere to
better than 15\%. We note in particular that the calculations reproduce
the observed decrease in TKE near the symmetric limit. This effect
is directly related to the extreme elongation ($Q_{20}=595\,\textrm{b}$)
reached by the nucleus at symmetric scission in Fig. \ref{cap:240pu-scl23-new}.

\begin{figure}
\includegraphics[%
  scale=0.44,
  angle=-90]{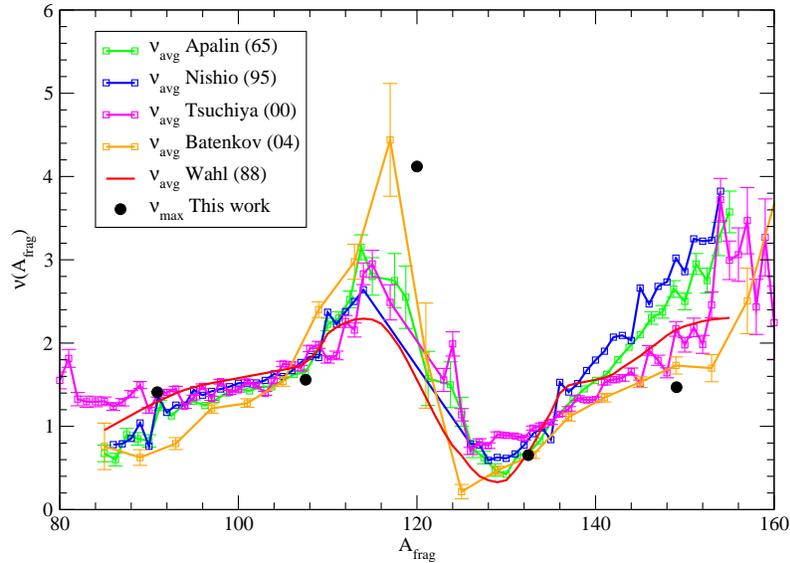}

\caption{\label{cap:nu-a}Comparison between measured \cite{nishio95,tsuchiya00,batenkov04,apalin65}
and calculated average neutron multiplicities, plotted as a function
of fragment mass.}
\end{figure}

In Fig. \ref{cap:nu-a}, we compare calculated and measured average
neutron multiplicities $\nu\left(A\right)$ as a function of fragment
mass. In the HFB calculations, the neutron multiplicities are obtained
using the very simplistic formula\begin{eqnarray*}
\nu\left(A\right) & = & \frac{E_{x}\left(A\right)}{B_{n}\left(A\right)+K_{n}}\end{eqnarray*}
where $E_{x}\left(A\right)$ is the excitation energy of the fragment,
$B_{n}\left(A\right)$ is the neutron separation energy, and $K_{n}$
is the average kinetic energy of the emitted neutron (taken as $K_{n}$=
2 MeV here). The excitation energy of each fragment was calculated
as the difference between HFB energies of the system at scission (obtained
in this work) and in its ground state (taken from the AMEDEE database
of ground-state HFB calculations \cite{amedee}). Because the two-body
center-of-mass contribution is calculated for the fissioning nucleus,
a correction (of order $\sim$ 3 MeV) was applied to the excitation
energy of individual fragments to restore the appropriate center-of-mass
energy for that fragment. The neutron-multiplicity data exhibit a
great deal of variability near symmetric fission, as well as for heavy
asymmetric fragments. Nevertheless, the calculations are in excellent
agreement with the data, except perhaps near the symmetric limit.
The most recent data, from Batenkov et al. \cite{batenkov04}, indicate
a sharp rise up to $\nu\approx4.4$ near $A$ = 117, not far from
the $\nu\approx4.1$ near $A$ = 120 found in the present calculation.
However, the remaining data \cite{nishio95,tsuchiya00,apalin65} and
Wahl evaluation \cite{wahl88} peak at a lower value of 2.3-3.2 near
$A$ = 114, albeit with significant experimental uncertainties. On
the other hand, the dip in multiplicity around the nearly-spherical
$A$ = 130 fragments is well reproduced by the calculations.

\section{Conclusion}

In this work, we have calculated fission-fragment properties for the
$^{239}\textrm{Pu}\left(\textrm{n}_{\textrm{th}},\textrm{f}\right)$
reaction in a fully microscopic approach using static constrained
HFB calculations with a finite-range interaction. Near scission, we
have separated the Slater-determinant HFB solution into an \emph{anti-symmetrized}
product of two distinct Slater determinants, one corresponding to
each fragment. We have introduced a quantitative criterion to identify
scission configurations based on building up a sufficient amount of
available energy in the fissioning system to overcome the attractive
part of the interaction between the fragments. Using this criterion,
we have calculated the kinetic and excitation energies of the fragments
and found them to be in very good agreement with experimental data.
In the future, we will extend the calculations of fragment properties
to more points along the scission line in Fig. \ref{cap:240pu-scl23-new}
and analyze the approach to scission as a function of the distance
between fragments. We will also explore the impact on these properties
of a full dynamical treatment of fission, using the approach developed
in \cite{berger84,goutte05}.

\section{Acknowledgments}

This work was performed under the auspices of the U.S. Department
of Energy by the Lawrence Livermore National Laboratory under Contract
DE-AC52-07NA27344.

\end{document}